\documentclass[11pt,a4paper]{article}
\usepackage{amsmath,amssymb}
\usepackage{epsfig,graphicx}

\begin{document}

\title{\bf Mass Spectrum, Actons and\\
Cosmological Landscape\footnote{Based on talks given  at
"Quarks-2006" St. Petersburg, Russia, 19-25 May, 2006; "Beyond the
Quantum", 29 May 2006 - 2 June 2006, Leiden Universitet, The
Netherlands;
 " Quantum Theory: Reconsideration of
Foundations-3", June 4- 9, 2006, Vaxjo Universitet, Sweden;
 IV Summer School in Modern Mathematical Physics,
 3 - 14 September 2006, Belgrade, Serbia.}
}
\author{V.~V.~Kozlov
and I.~V.~Volovich
\\
\small{\em Steklov Mathematical Institute}
\\
\small{\em Gubkin St. 8, 119991, Moscow, Russia}
}
\date{}
\maketitle
\begin{abstract}
It is suggested that the properties of the mass spectrum of
elementary particles could be related with cosmology. Solutions of
the Klein-Gordon equation on the Friedmann type manifold with the
finite action are constructed. These solutions (actons) have a
discrete mass spectrum. We suggest that such solutions could select
a universe from  cosmological landscape. In particular the solutions
with the finite action on de Sitter space are investigated.
\end{abstract}

\section{Introduction}

Understanding of the mass spectrum of elementary particles is an outstanding
problem for  physics. Why the elementary particles  have their
observed pattern of masses? Note that there is no answer to even a simpler question why
the mass spectrum is discrete.

The  mass in quantum field theory \cite{BS} is considered as an
arbitrary parameter but  in nature there is only a discrete set
of masses of elementary particles.

We investigate the following proposal. We suggest that {\it the mass
parameter in the Klein-Gordon equation should be such that there
exists a corresponding classical solution with the finite action}.
In other words  we consider an eigenvalue problem for the
Klein-Gordon equation.

There are not such nontrivial solutions in the Minkowski space.
However we will show that  the square integrable solutions of the
Klein-Gordon equation on an important class of manifolds  do exist.
Moreover such solutions have  the finite action (we call them {\it
actons}) and they exist  only for some discrete values of the mass,
i.e. we obtain quantization of masses.

A finite mass spectrum was first obtained in \cite{Koz} for  de
Sitter space, solutions on Lorentzian manifolds are considered in
\cite{KV}.

The requirement of the finiteness of the action in the
Lorentz signature is natural, for example, in the case when we study
the wave function of the Universe in real time in the
semiclassical approximation \cite{Vil} where it is just
$ \exp(iS),~S$ being the action. It is  known that there are
solutions of some nonlinear equations with finite action
(instantons) but they exist only in the Euclidean time.

In quantum gravity and string landscape picture (see \cite{Sakh,ADV,
Wei,CD,Tye} and refs therein) our observed universe is viewed within
a multiverse which contains every possible type of vacuum and the
law of physics are determined by the anthropic principle. We suggest
that {\it the principle of the finiteness of the action} proposed in
this work  can be used to restrict the multiplicity of universes.

The actons are classical solutions.
They define masses of elementary particles.
Then we have to quantize
the system in this background as we do with solitons and instantons.

The values of masses of scalar particles obtained this way so far
are not very realistic but the mechanism of how to get the discrete mass spectrum
seems does work.

The idea that the boundedness of the mass spectrum might be related
with de Sitter geometry in the momentum space is considered in
\cite{Kad}. A symmetry which exploits the feature that de Sitter and
Anti de Sitter space are related by analytic continuation is
considered in \cite{HN}.

Let  $M$ be an $(n+1)$-dimensional manifold with a Lorentz metric
$g_{\mu\nu},~\mu,\nu=0,1,...,n$. Consider the Klein-Gordon
equation \cite{BD} on $M$ for the real valued function
$f$
\begin{equation}
\label{1} \square f+\lambda f=0.
\end{equation}
Here
$$
\square f=\nabla_{\mu}\nabla^{\mu}f=
\frac{1}{\sqrt{|g|}}\partial_{\mu}(\sqrt{|g|}g^{\mu\nu}\partial_{\nu}f),
$$
$g$ is the determinant of $(g_{\mu\nu})$ and the real parameter
$\lambda$ corresponds to the mass square.

We are interested in deriving the values of $\lambda$ for which
there exist classical solutions $f\in C^2(M),$ satisfying the
condition
\begin{equation}
\label{2}
 \int_{M}f^2\sqrt{|g|}dx<\infty
\end{equation}
The condition (\ref{2}) was first considered in \cite{Koz} for
solutions of the Klein-Gordon equation on de Sitter space. Let
us note that the condition (\ref{2}) includes the integration not
only over the spatial variables as it is done usually for the
quantum Klein-Gordon field \cite{BD} but also over the time-like
variable.

The paper is composed as follows. In the next section square
integrable solutions of the rather general type of manifolds which
are called the Friedmann type manifolds are considered. Then
solutions with the finite action (actons) on de Sitter space and on
the Friedmann space are considered.

\section{ Solutions on  the Friedmann type manifolds}

Let us consider a manifold $M=I\times N^n$ with a metric:
\begin{equation}
\label{1A}
 ds^2=g_{\mu\nu}dx^{\mu}dx^{\nu}=dt^{2} -a^2(t)dl^2.
\end{equation}
Here  $I$ is an interval on the real axis, $I\subset \mathbb{R}$,
$~$ $a(t)$ is a smooth positive function on $I$, $~$ $N^n$ a
Riemannian manifold and
\begin{equation}
\label{1B} dl^2=h_{ij}(y)dy^i dy^j,~~i,j=1,...,n
\end{equation}
is a Riemannian metric on $N^n.$ Such manifolds $(M,g_{\mu\nu})$
will be called the Friedmann type manifolds.

Eq. (\ref{1}) for the metric (\ref{1A})takes the form
\begin{equation}
\label{3}
 \ddot{f}+\frac{n}{a}\dot{a}\dot{f}-\frac{1}{a^2}\Delta_h
f+\lambda f=0
\end{equation}
where $\Delta_h$  is the Laplace-Beltrami operator fot the
metric $h_{ij}$,
\begin{equation}
\label{4} \Delta_h
f=\frac{1}{\sqrt{h}}\partial_{i}(\sqrt{h}h^{ij}\partial_{j}f)
 \end{equation}
and the condition (\ref{2}) reads
\begin{equation}
\label{2c}
 \int_{M}f^2\sqrt{|g|}dx=\int_{I\times
 N^n}f^2|a|^n\sqrt{h}dtdy<\infty
\end{equation}

Let $q\geq 0$ be the eigenvalue of the operator $-\Delta_h$ on
$N^n$ and $\Phi=\Phi(y)$ is the corresponding eigenfunction:
\begin{equation}
\label{5} -\Delta_h\Phi=q\Phi,
 \end{equation}

\begin{equation}
\label{6} \int_{N^n}\Phi^2\sqrt{h}dy<\infty
 \end{equation}
We set
\begin{equation}
\label{7} f=B(t)a(t)^{-\frac{n}{2}}\Phi(y).
 \end{equation}
 Then from (\ref{3}),(\ref{5}) we obtain the Sturm-Liouville (Schrodinger)
 equation
\begin{equation}
\label{9w} \ddot{B}+[\lambda - v(t) ]B=0
 \end{equation}
 where
\begin{equation}
\label{6AB}
 v(t)=\frac{n}{2}\frac{\ddot{a}}{a}+\frac{n}{2}(\frac{n}{2}-1)
 \frac{\dot{a}^2}{a^2}-\frac{q}{a^2}
 \end{equation}
We look for solutions $B(t)$ of Eq. (\ref{9w}) in $ L^2(I)$ since
for functions of the form (\ref{7}) the condition (\ref{2c}) takes
the form

$$
\int_{I}B^2 dt<\infty.
$$
Consider the case  $I= \mathbb{R}.$

 {\it Let $M=\mathbb{R}\times N^n$ be the Friedman
type manifold with the metric of the form (\ref{1A}),(\ref{1B})
such that there exists a solution $\Phi$ of Eq. (\ref{5}) on $N^n$
which is not identically vanishing with an eigenvalue $q\geq 0$.
Let the smooth positive function  $a(t)$ on $\mathbb{R}$ is such
that $v(t)$ (\ref{6AB}) satisfies the condition
\begin{equation}
\label{6ABC}
 v(t)\rightarrow\infty ~~if ~~|t|\rightarrow\infty.
 \end{equation}

  Then for given $q,\Phi$
 (\ref{5}),(\ref{6}), the problem (\ref{1}),(\ref{2}) has an infinite family of
 solutions (actons) $f_j=B_j(t)a(t)^{-\frac{n}{2}}\Phi(y)$ with eigenvalues
 $\lambda_j,~j=1,2,...$  and moreover
 $\lambda_j\rightarrow\infty ~ when ~ j\rightarrow\infty.$ }

The statement is evident from quantum mechanics.
It follows from a well known result by Weyl and Titchmarsh (see
for example \cite{Tit}, Sect.5.12-5.13) that under condition
(\ref{6ABC}) the Sturm-Liouville problem (\ref{9w}) has a
discrete spectrum in $ L^2(\mathbb{R})$.

  Example. Let us take
\begin{equation}
\label{10} a(t)=C\exp(\alpha t^{2k}),~C>0,~\alpha >0,~k>1.
 \end{equation}
 Then
 $$
 v(t)=\frac{n}{2}[\alpha 2k(2k-1)t^{2k-2}+
 (\frac{n}{2}-1)(\alpha 2k)^2t^{4k-2}]-qC^{-2}\exp(-\alpha t^{2k})
 $$
In this case the condition (\ref{6ABC}) is satisfied. One has a
discrete spectrum $\lambda_1,\lambda_2,...$ and
 $\lambda_j\rightarrow\infty $~ when~ $j\rightarrow\infty.$

\section{Solutions on de Sitter space}

 For de Sitter space one has:
 $M=\mathbb{R}\times S^3$,
\begin{equation}
\label{9рp} ds^2=dt^{2} - \cosh^{2} t \cdot h_{ij}(y)dy^i dy^j,
 \end{equation}
 where $h_{ij}$ is the standard metric on the 3-dimensional sphere
  $S^3$. The eigenvalues of the operator  $-\Delta_h$ on the 3-sphere
  are equal to
 $q=j(j+2),~j=0,1,2,...$ and
\begin{equation}
\label{11a} v(t)=\frac{9}{4}-[\frac{3}{4}+j(j+2)]\frac{1}{\cosh^2
t}
 \end{equation}
 We set
\begin{equation}
\label{11G}
 \alpha=\frac{3}{4}+j(j+2), ~\nu^2=\frac{9}{4}-\lambda
 \end{equation}
Then Eq. (\ref{9w}) takes the form
\begin{equation}
\label{9r} \ddot{B}+[\frac{\alpha}{\cosh^2 t} -\nu^2 ]B=0
 \end{equation}

Theory of Eq. (\ref{9r}) is well known   \cite{Tit,Flu} and it was
used in \cite{Koz}
 to construct square integrable solution of the
 Klein-Gordon equation on de sitter space. Spectrum for positive values
 of  $\nu^2$ is discrete and for negative is continuous. We
 consider the first case, $\nu^2>0$.

 Eq.
(\ref{9r}) for $\alpha >0$ has a solution in $L^2(\mathbb{R})$ iff
\begin{equation}
\label{9р} 0<\nu=\frac{1}{2}(\sqrt{1+4\alpha}-1)-n, ~ n=0,1,2,...
 \end{equation}
 In our case, due to (\ref{11G}),
 $$
 0<\nu=j+\frac{1}{2}-n, ~~ j,n=0,1,2,...
 $$
There is a family of square integrable solutions of
 Eq. (\ref{9w}) with eigenvalues $\lambda$ of the form
\begin{equation}
\label{13q}
\lambda_{jn}=\frac{9}{4}-(j+\frac{1}{2}-n)^2,~~(j,n=0,1,2,...,j+\frac{1}{2}-n>0)
 \end{equation}
If $\lambda_{jn} \geq 0$ then we should have
 $$
0<j+\frac{1}{2}-n\leq\frac{3}{2},~~ j,n=0,1,2,...
 $$
and therefore either $j=n$ and $\lambda_{jn}=2$, or $j=n+1$ and
$\lambda_{jn}=0$.

In the case $j=n,~ \nu=1/2$ for any $j=0,1,2,...$  Eq. (\ref{9r})
has a solution (acton) in  $L^2(\mathbb{R})$ of the form
 \begin{equation}
\label{14V} B_j(t)=\frac{1}{(\cosh t)^{1/2}}\sum_{s=0}^j \frac
{(-j)_s(j+2)_s}{(3/2)_s s!}\frac{1}{(e^{2t}+1)^s},
 \end{equation}
 $$
 (k)_0=1,~(k)_s=k(k+1)...(k+s-1).
 $$
In the case $j=n+1,~ \nu=3/2$  for any $j=1,2,...$ Eq. (\ref{9r})
has a solution in  $L^2(\mathbb{R})$  of the form
 \begin{equation}
\label{14VV} B_j(t)=\frac{1}{(\cosh t)^{3/2}}\sum_{s=0}^{j-1}
\frac {(1-j)_s(j+3)_s}{(5/2)_s s!}\frac{1}{(e^{2t}+1)^s}
 \end{equation}
We have obtained that if the eigenvalues $\lambda =\lambda_{jn}\geq 0$
 then either $\lambda =0,$ or $\lambda =2$.

\section{Solutions on the Friedmann space}

 {\bf 1.} In the inflation cosmology the following form of the Friedmann-de Sitter
 metric is often used:
\begin{equation}
\label{13} ds^2=dt^2-e^{2Ht} \cdot h_{ij}(y)dy^i dy^j,
 \end{equation}
где $h_{ij}$ is a Riemannian metric on a compact 3-dimensional
manifold,
 $0<t<\infty$ and $0<H$ is  Hubble`s constant. In this case the
 function $v(t)$ (\ref{6AB}) is
 \begin{equation}
\label{14} v(t)=\frac{9}{4}H^2-qe^{-2Ht}
 \end{equation}
Eq. (\ref{9w}) on the semi-axis with boundary conditions
$B(0)=B(\infty)=0$ has an eigenvalue  in this case. If the
parameter $t$ is interpreted as the radius in spherical
coordinates then we get the known model of deuteron (see, for
example \cite{Flu}). The solution has the form
$$
B(t)=J_{\nu}(ce^{-Ht}),
$$
where
$$
c=\frac{\sqrt{q}}{H}>0,~~\nu=\frac{\sqrt{9H^2-4\lambda}}{2H}>0,
$$
and $J_{\nu}$ is the Bessel function. The eigenvalue $\lambda$ is
derived from the relation $J_{\nu}(c)=0.$

 {\bf 2.} The Friedmann metric has the form
\begin{equation}
\label{13N} ds^2=dt^2-a^2(t) h_{ij}(y)dy^i dy^j
 \end{equation}
where $h_{ij}$ is a Riemannian metric on the manifold of positive,
negative or flat curvature.  The function $a(t)$ is derived from the
Einstein-Friedmann equations
\begin{equation}
\label{13L} 3\dot{a}^2/a^2=8\pi \rho-3k/a^2,~~3\ddot{a}/a=
-4\pi(\rho+3p),
 \end{equation}
where $k=1, -1, 0$ for the manifolds for manifolds of the positive,
negative, or flat curvature respectively. The pressure $p$ and the
mass density $\rho$ are related by an equation of state $p=p(\rho)$.
In particular, for massless thermal radiation $(p=\rho/3)$ in a
3-dimensional torus $(k=0)$ one has
\begin{equation}
\label{13T} a(t)=c\sqrt{t},~~c>0,~~ 0<t<\infty.
 \end{equation}
In this case
\begin{equation}
\label{13TA} v(t)=-\frac{3}{16t^2}-\frac{q}{c^2t},~~q>0
 \end{equation}
and  the Sturm-Liouville equation (\ref{9w}) has a discrete
spectrum for negative $\lambda$ :
\begin{equation}
\label{13TB} \lambda_n=-\frac{4q^2}{c^4(4n+1)^2},~~n=1,2,...
 \end{equation}
Indeed, if we denote $\lambda =-\nu^2,~\nu >0$ and define a new
function $\varphi (x)$ by
$$
B(t)=e^{-\nu t}t^{1/4}\varphi(2\nu t)
$$
then from the Sturm-Liouville equation
$$
\ddot{B}(t)+[\lambda +\frac{3}{16t^2}+\frac{q}{c^2t} ]B(t)=0
$$
we obtain that the function $\varphi (x)$ satisfies the equation
for the degenerate hypergeometric function
$$
x\varphi^{\prime\prime}(x)+(\frac{1}{2}-x)\varphi^{\prime}(x)
-(\frac{1}{4}-\frac{q}{2\nu c^2})\varphi(x)=0.
$$
It is known that the last equation has  solutions with the required
behavior at infinity
 only if
$$
\frac{1}{4}-\frac{q}{2\nu c^2}=-n,~~n=1,2,...
$$
which leads to (\ref{13TA}).

 For a compact 3-dimensional manifold of negative curvature
$(k=-1)$ the function $a(t)$ has the form
\begin{equation} \label{13LL} a(t)=\sqrt{t^2-c^2},~~0<c<t<\infty
 \end{equation}
and the corresponding  Sturm-Liouville problem also has a
discrete spectrum for negative $\lambda$ .

\section{Discussions and Conclusions  }

{\bf 1}. An action for the Klein-Gordon equation (\ref{1}) has
the form
 \begin{equation}
\label{15} S=\frac{1}{2}\int_M[(\nabla f,\nabla f)-\lambda
f^2]\sqrt{|g|}dx
 \end{equation}
where
$$
(\nabla f,\nabla f)=g^{\mu\nu}\partial_{\mu}f\partial_{\nu}f.
$$
On solutions of the form  (\ref{7}) on the Friedmann type manifolds
the action takes the form
 \begin{equation}
\label{15B}  S=\frac{1}{2}\int_{I\times
 N^n}[(\dot{B}-\frac{n}{2}
 \frac{\dot{a}}{a}B)^2\Phi^2-a^{-2}B^2h^{ij}
 \partial_i\Phi\partial_j\Phi -\lambda B^2\Phi^2]dt\sqrt{h}dy
 \end{equation}
On the solutions on de Sitter space of the form (\ref{14V}),
(\ref{14VV}) the integral (\ref{15B}) is convergent, i.e. the
action  is finite (moreover, $S=0$).

 {\bf 2.} Let us make the substitution into Eq.  (\ref{3})
$$
f=u(y,t)a(t)^{-\frac{n}{2}}
$$
Then we obtain the following equation
 \begin{equation}
\label{16} \ddot{u}-a(t)^{-2}\Delta_h u +[\lambda - w(t)]u=0
 \end{equation}
where
$$
w(t)=\frac{n}{2}\frac{\ddot{a}}{a}+\frac{n}{2}(\frac{n}{2}-1)
 \frac{\dot{a}^2}{a^2}
$$
We look for solutions satisfying the condition
$$
\int_{\mathbb{R}\times
 N^n}u(y,t)^2dt\sqrt{h}dy<\infty
$$
There exists a well developed spectral theory for elliptic
differential operators (see, for example \cite{Tit, DS}.
 There is a spectral theory of the Liouville operator in ergodic theory
 of dynamical systems \cite{AKN }. It would
be interesting to develop a spectral theory for hyperbolic
equations.

 Consider for example on the Schwartz space $S(\mathbb{R}^2)$
 of the functions $u=u(x,t)$ on the plane the hyperbolic
 differential operator of the form
\begin{equation}
\label{17} Au=\frac{\partial^2}{\partial
t^2}u-\frac{\partial^2}{\partial x^2} u +\phi(x,t)u
 \end{equation}
where the smooth real valued function $\phi(x,t)$ admits a power
bound  on the variables $x,t$ at infinity. The operator $A$ admits a
self-adjoint extension in $L^2(\mathbb{R}^2)$. In the particular
simple case when the function $\phi$ has the form $\phi=x^2-t^2,$
the operator   is the difference of the Schrodinger operators for
two harmonic oscillators (this case was disscussed by Ginzburg,
Manko and Markov \cite{GM }). Hence it has in $L^2(\mathbb{R}^2)$ a
complete system of eigenfunctions
\begin{equation}
\label{17R} u_{jn}=H_j(t)H_n(x)\exp\{-\frac{1}{2}(t^2+x^2)\},
 \end{equation}

$$
Au_{jn}=
 \lambda_{jn}u_{jn},~~j,n=0,1,2,...,
$$
with the corresponding eigenvalues $\lambda_{jn}=2(n-j).$ Here $H_j$
are the Hermite polynomials.

The action
$$
S=\frac{1}{2}\int_{\mathbb{R}^2}(\dot{u}^2-u_x^2-\phi u^2+\lambda
u^2)dtdx
$$
is finite on solutions (\ref{17R}).

For a recent discussion of  field theories with unbounded energy and
its cosmological applications see \cite{AV2}.

 {\bf 3.}  Note that together with  (\ref{1}) the so called
 equation with conformal coupling is considered:
\begin{equation}
\label{18} \square f+\xi Rf+\lambda f=0.
 \end{equation}
Here $R$ is the scalar curvature of the manifold $M$ and $ \xi
=(n-1)/4n.$  For de Sitter space $R=12,~ \xi=1/6$. In this case
Eq.  (\ref{18}) $\square f+(2+\lambda )f=0$ has square integrable
solutions if $2+\lambda =\lambda_{jn}$ (\ref{13q}) and in
particular if  $\lambda=0$.

To conclude, in this note the square integrable solutions with the
finite action (actons) of the Klein-Gordon equation for scalar field
on manifolds have been considered. It would be interesting to study
solutions with the finite action on more general manifolds and also
such solutions for equations for fields with higher spins.


\section*{Acknowledgements}

V.K. is supported in part by RFBR grant  05-01-01058 and I.V. by RFBR
05-01-00884, NS-6705.2006 and INTAS 03-51-6346.

\end{document}